\documentclass[a4paper,aps,pra,showpacs,twocolumn,superscriptaddress]{revtex4-1}			
   
\usepackage[utf8]{inputenc}  
\usepackage[T1]{fontenc}     
\usepackage[british]{babel}  
\usepackage{lmodern}  
\usepackage[usenames,dvipsnames]{color} 
\usepackage[colorlinks,citecolor=blue,linkcolor=magenta,urlcolor=blue]{hyperref}  
\usepackage{graphicx}
\usepackage{tikz}
\usepackage[babel]{microtype}  
\usepackage{amsmath,amssymb,amsthm,bm,mathtools,amsfonts,mathrsfs,bbm,dsfont} 
\usepackage{xspace}  

\usepackage{physics}
\newcommand{\id}{\ensuremath{\mathds{1}}}
\usepackage{bbold}

\newcommand{\meas}[1]{\mathscr{#1}}
\newcommand{\ent}[1]{\mathbf{#1}}
\renewcommand{\norm}[1]{\lVert #1 \rVert}

\newtheoremstyle{mystyle}
  {6pt}
  {6pt}
  {\normalfont}
  {0pt}
  {\bf}
  {.}
  { }
  {}

\theoremstyle{mystyle}
\newtheorem{theorem}{Theorem}

\newtheorem{definition}[theorem]{Definition}

\newtheorem{observation}[theorem]{Observation}

\begin{document}
\nonfrenchspacing
\title{Monogamy relations of quantum coherence between multiple subspaces}

\begin{abstract}
Quantum coherence plays an important role in quantum information protocols that provide an advantage over classical information processing. The amount of coherence that can exist between two orthogonal subspaces is limited by the positivity constraint on the density matrix. On the level of multipartite systems, this gives rise to what is known as monogamy of entanglement. On the level of single systems this leads to a bound, and hence, a trade-off in coherence that can exist between different orthogonal subspaces. In this work we derive trade-off relations for the amount of coherence that can be shared between a given subspace and all other subspaces based on trace norm, Hilbert-Schmidt norm and von Neumann relative entropy. From this we derive criteria detecting genuine multisubspace coherence.
\end{abstract}

\author{Tristan Kraft}
\email{tristan.kraft@uni-siegen.de}
\affiliation{Naturwissenschaftlich-Technische Fakult\"at, Universit\"at Siegen, Walter-Flex-Str. 3, D-57068 Siegen, Germany}
\affiliation{SUPA and Department of Physics, University of Strathclyde, G40NG Glasgow,
United Kingdom}
\author{Marco Piani}
\affiliation{SUPA and Department of Physics, University of Strathclyde, G40NG Glasgow,
United Kingdom}

\date{\today}  

\maketitle

\section{Introduction}
Quantum coherence refers to the possibility of preparing a system in a state of superposition, a feature that distinguishes quantum mechanics from classical physics. Recently, a lot of effort has been devoted to construct a resource theory of coherence, and to formalize the quantification and manipulation of coherence~\cite{aberg2006, baumgratz2014, streltsov2017a}. Furthermore, coherence in multipartite systems has been studied~\cite{yao2015, radhakrishnan2016, streltsov2017b, Kraft2018} and a connection between the resource theory of coherence and entanglement has been established~\cite{streltsov2015, streltsov2016, chitambar2016}.

An interesting feature of entanglement is the fact that it cannot be distributed arbitrarily amongst many parties. For instance, if two parties share a maximally entangled state, no entanglement can exist between one of those parties and a third one. This is referred to as monogamy of entanglement. It was first made precise by Coffman, Kundu and Wootters in Ref.~\cite{Coffman2000}, where they showed that for the case of three-qubit pure states it holds $\mathcal{C}_{A|B}+\mathcal{C}_{A|C}\leq\mathcal{C}_{A|BC}$, where $\mathcal{C}_{A|B}$ is the entanglement quantifier known as concurrence. The concurrence of a two-qubit state $\varrho_{AB}$ can be calculated as $\mathcal{C}_{A B}=\max \left\{0, \lambda_{1}-\lambda_{2}-\lambda_{3}-\lambda_{4}\right\}$, where the $\lambda_i$ are the increasingly ordered eigenvalues of the hermitian operator $\sqrt{\sqrt{\rho} \tilde{\rho} \sqrt{\rho}}$ and $\tilde{\rho}=\left(\sigma_{y} \otimes \sigma_{y}\right) \rho^{*}\left(\sigma_{y} \otimes \sigma_{y}\right)$. This result can be extended to three-qubit mixed states~\cite{Coffman2000} and was later generalized to the case of arbitrarily many qubits by Osborne and Verstraete in Ref.~\cite{Osborne2006}.

The constraint that leads to such a bound is the positivity of the quantum state, i.e. a valid quantum state is described by a positive semidefinite operator with trace one. The positivity of the quantum state limits the amount of coherence that can be shared between any two orthogonal subspaces. On the level of multipartite systems this leads to monogamy of coherence that can be shared between multiple systems \cite{radhakrishnan2016, kumar2017} and, thus, limits the amount of entanglement that can be shared between multiple parties.

The fact that coherence cannot be shared arbitrarily leads to limitations on how well quantum states can be distinguished when the measurements that are being performed cannot access all subspaces. Consider a state $\varrho(0)$ that is subjected to unitary time evolution by the Hamiltonian $H=\sum_i E_iP_i$, where $P_i$ denotes the projector onto the eigenspace corresponding to the energy $E_i$. If the initial state exhibits some coherence between the energy eigenspaces, it will start to evolve in a state that is distinguishable from the initial state. The maximum distinguishability, using measurements that are limited to the subspace $\mathcal{H}_{E_i}\oplus \mathcal{H}_{E_j}$, depends on the amount of coherence that is shared between those subspaces, quantified by the trace norm of the corresponding off-diagonal block of the density matrix.

In this manuscript we investigate the constraints imposed by positivity on the structure of density matrices. We will derive a trade-off in the coherence that can be shared between one subspace and all other orthogonal subspaces. In Section~\ref{sec:2} of this paper we will set the scenario and recall some important results about Schatten norms. In Section~\ref{sec:3} we will state the main results, namely we will derive bounds on the amount of coherence that can be shared between one and all the other orthogonal subspaces, based on trace norm, Hilbert-Schmidt norm and the von Neumann relative entropy. In Section~\ref{sec:4} we will illustrate our results by applying them to the case of a single qutrit. In Section~\ref{sec:5} we will show how our approach can be used to detect genuine multisubspace coherence.

\section{Preliminaries}\label{sec:2}

Consider a finite-dimensional Hilbert space $\mathcal{H}$. We denote by $\textbf{P}=\qty{P_i}_{i=0}^N$ a complete set of orthogonal projectors, each acting like the identity on the orthogonal subspaces $\mathcal{H}_i$, $\bigoplus_{i=0}^N \mathcal{H}_i = \mathcal{H}$. Any state $\varrho$ can decomposed into its block components with respect to $\textbf{P}$: $\varrho = \qty[\varrho_{ij}]_{i,j=0}^{N}=\qty[P_i\varrho P_j]_{i,j=0}^{N}$. For the diagonal blocks it holds that $\varrho_{ii}^{\dagger}=\varrho_{ii}$ and for the off-diagonal blocks we have that $\varrho_{ij}^{\dagger}=\varrho_{ji}$ for $i\neq j$. Whenever we consider the components of the state $\varrho$ with respect to two specific subspaces, we adopt the following shorthand notation
\begin{equation*}
    \varrho^{(ij)} = \begin{bmatrix} 
    \varrho_{ii} & \varrho_{ij} \\
    \varrho_{ij}^{\dagger} & \varrho_{jj}
    \end{bmatrix}.
\end{equation*}
It is worth noting how this relates to the theory of block-coherence. This was studied in Ref.~\cite{aberg2006} and more recently in Refs.~\cite{bischof2019a, bischof2019b}.
A state is called block incoherent if it can be written as \begin{equation}
    \varrho=\text{diag}(\varrho_{ii}),
\end{equation}
otherwise the state is deemed to be coherent with respect to $\textbf{P}$.

The coherence that exists between two different subspaces $i$ and $j$
is encoded in the off-diagonal block $\varrho_{ij}$. The amount of coherence between these blocks does not depend on the choice of basis within these blocks, that is, its amount should be invariant under block-diagonal unitary transformations of the form $U=\bigoplus_{i}U_i$. Such a unitary transformation changes the off-diagonal blocks to $\varrho_{ij}\mapsto U_i^{\dagger}\varrho_{ij} U_j$. Hence, to quantify the amount of coherence that is shared between different subspaces we take a unitarily invariant matrix norm of the off-diagonal blocks, such as the Schatten $p$-norms. Given a matrix $M\in \mathcal{M}_{n,m}$ the Schatten $p$-norms are defined on the vector of singular values $\Vec{\sigma}(M)$ of $M$ by
\begin{equation*}
\norm{M}_p^p=\left( \sum_{i=1}^{\min\qty{m,n}}\sigma_i(M)^p\right)^{1/p}.  
\end{equation*}
Here $p\in[1,\infty]$ and the singular values are non-negative and assumed to be ordered such that $\sigma_1\geq \sigma_2 \geq \dots \geq \sigma_{\min\{m,n\}}\geq 0$. They are the eigenvalues of the Hermitian operator $\abs{M}=\sqrt{M^{\dagger} M}$ and thus the Schatten $p$-norms can be expressed as $\norm{M}_p^p=\tr(M^{\dagger}M)^{\frac{p}{2}}$. For $p=1$ the norm $\norm{\cdot}_1=\norm{\cdot}_{tr}$ is called the trace norm, for $p=2$ the norm $\norm{\cdot}_2=\norm{\cdot}_{HS}$ is called the Hilbert-Schmidt norm and for $p=\infty$ (defined via a limit procedure) the norm $\norm{\cdot}_{\infty}$ is called the operator norm, and equal to the largest singular value of $M$. One of the main properties of Schatten norms is their isometric invariance, i.e. $\norm{UMV}_p=\norm{M}_p$ for all isometries $U,V$.

\section{Results}\label{sec:3}
Our aim is to capture the trade-off in coherence that can be shared between one specific subspace (without loss of generality, the first one, $\mathcal{H}_0$) and and all the other $N$ subspaces. First, we quantify the amount of coherence by taking the sum over the trace-norms and Hilbert-Schmidt norms of all the blocks that contain information about the coherence between the first and all other $N$ subspaces. Then, we will derive trade-off relations in terms of the von Neumann relative entropy.
\subsection{Trace norm}
Let us start by defining the following quantity
\begin{definition}
Given a set of projectors $\textbf{P}$ on $\mathcal{H}$ and a state $\varrho = \qty[\varrho_{ij}]_{i,j=0}^{N}$, the amount of coherence that is shared between the first and all other $N$ subspaces is quantified by
\begin{equation*}
\meas{C}^{tr}_{0:1\dots N}(\varrho)=\sum_{j=1}^N \norm{\varrho_{0j}}_{tr}.
\end{equation*}
\end{definition}
Let us derive an upper bound on this sum which results in a trade-off relation between the shared coherence. For a positive semidefinite matrix $M\geq 0$ which is of the form
\begin{equation}
\label{eq:blockmatrix}
M = \mqty(A&X\\X^{\dagger}&B),
\end{equation}
with $A,B\geq 0$, the off-diagonal blocks can always be written as
\begin{equation}
\label{eq:contraction}
    X=A^{1/2}KB^{1/2},
\end{equation}
where $K$ is a contraction, i.e. $K^{\dagger}K\leq \id$ \cite{bhatia2013}. It was shown in Ref.~\cite{horn1990} that for matrices of the form of Eq.~\eqref{eq:blockmatrix}
\begin{equation}
\label{eq:CSinequality}
    \norm{\abs{X}^q}^2 \leq \norm{A^q}\norm{B^q}
\end{equation}
holds for all unitarily invariant norms $\norm{\cdot}$ and all $q>1$. In particular, this inequality holds for the trace norm. By choosing $q=1$ we obtain
\begin{eqnarray*}
\meas{C}^{tr}_{0:1\dots N}(\varrho)=\sum_{k=1}^N \norm{\varrho_{0k}}_{tr} &\leq& \sum_{k=1}^N \sqrt{\norm{\varrho_{00}}_{tr}}\sqrt{\norm{\varrho_{kk}}_{tr}}\notag \\
&=&\sqrt{\tr\varrho_{00}}\sum_{k=1}^N \sqrt{\tr\varrho_{kk}}.
\end{eqnarray*}
Then, by the inequality between the arithmetic and quadratic mean, we obtain
\begin{eqnarray*}
\sqrt{\tr\varrho_{00}} \sum_{k=1}^N \sqrt{\tr\varrho_{kk}} &\leq& \sqrt{N} \sqrt{\tr\varrho_{00}} \sqrt{\sum_{k=1}^N \tr\varrho_{kk}}\notag \\
&=& \sqrt{N}\sqrt{\tr\varrho_{00}(1-\tr\varrho_{00})}
\end{eqnarray*}
\begin{observation}
\label{obs:bound1norm}
We find that the coherence between the first and all other $N$ subspaces is bounded by
\begin{equation}
\label{eq:bound1norm}
    \meas{C}_{0:1\dots N}^{tr}(\varrho)=\sum_{k=1}^N \norm{\varrho_{0k}}_{tr}\leq \sqrt{N}\sqrt{\tr\varrho_{00}(1-\tr\varrho_{00})}.
\end{equation}
\end{observation}
Let us discuss this result. First, this bound leads to a trade-off in coherence between the subspaces and the bound only depends on the number of subspaces involved and the accumulated probability in the first block. Second, the bound provided in Eq.~\eqref{eq:bound1norm} is tight, meaning that there always exists a state saturating the inequality. Consider a block $\varrho_{00}$, we can define the state
\begin{equation}
    \label{eq:saturatingstate}
   \sigma = \ketbra{\varphi}{\varphi} \otimes \Hat{\varrho}_{00},
\end{equation}
with $\ket{\varphi}=\qty(\sqrt{\tr\varrho_{00}}, \sqrt{(1-\tr\varrho_{00})/N},\dots ,\sqrt{(1-\tr\varrho_{00})/N})$ and $\Hat{\varrho}_{00}=\varrho_{00}/\tr\varrho_{00}$. It is straightforward to show that this state saturates the bound. Note that if all the blocks are one dimensional, $\varrho_{00}$ is a probability and hence the saturating state is pure. But this is not true in general. Furthermore, not every pure state saturates the bound.

The bound in Observation~\ref{obs:bound1norm} still depends on the number $N$ of subspaces. Next we consider a slight variation in the quantifier, and derive a bound that is quadratic in the trace norm of the off-diagonal blocks. A calculation similar to the previous one leads to the following.
\begin{observation}
\label{ob:quadratic_quantifier}
Given a state $\varrho = \qty[\varrho_{ij}]_{i,j=0}^{N}$ using Eq.~\eqref{eq:CSinequality} we obtain the following bound
\begin{eqnarray}
\label{eq:qudratic}
    \meas{C}_{0:1\dots N}^{tr,2}(\varrho) &\equiv& \sum_{i=1}^N \norm{\varrho_{0i}}_{tr}^2 \notag\\
    &\leq& \sum_{i=1}^N \norm{\varrho_{00}}_{tr} \norm{\varrho_{ii}}_{tr}\notag \\
    &=& \norm{\varrho_{00}}_{tr} \sum_{i=1}^N \norm{\varrho_{ii}}_{tr}\notag \\
    &=& \tr[\varrho_{00}](1-\tr[\varrho_{00}]). \notag
\end{eqnarray}
\end{observation}
Again, this bound is tight in a sense that there always exist a state that saturates the bound. Specifically, it is straightforward to see that the state $\sigma = \ketbra{\varphi}{\varphi}\otimes \hat{\varrho}_{00}$, with $\hat{\varrho}_{00}=\varrho_{00}/\tr\varrho_{00}$ and $\ket{\varphi}=\qty(\sqrt{\tr\varrho_{00}}, \sqrt{\tr\varrho_{11}},\dots,\sqrt{\tr\varrho_{NN}})$, saturates the bound. In general, a pure state $\ket{\psi}$ always saturate the inequality, independently of the dimensions of the subspaces $\mathcal{H}_i$. Indeed, for a pure state $\ket{\psi}$, one has $\|\rho_{0i}\|^2_{tr} = \|P_0 \ketbra{\psi}{\psi}P_i\|^2_{tr} = \|P_0 \ket{\psi} \|^2_\infty \|P_i \ket{\psi} \|^2_\infty = \|P_0 \ketbra{\psi} P_0\|_{tr} \|P_i \ketbra{\psi} P_i\|_{tr} = \|\rho_{00}\|_{tr} \|\rho_{ii}\|_{tr}$.
In particular we observe that in general Observation~\ref{obs:bound1norm} follows from Observation~\ref{ob:quadratic_quantifier}. To see this, take Eq.~\eqref{eq:qudratic} and use the inequality between  arithmetic and quadratic mean. One obtains $1/N\qty(\sum_{i=1}^N \norm{\varrho_{0i}}_{tr})^2\leq\sum_{i=1}^N \norm{\varrho_{0i}}_{tr}^2$,from which Observation~\ref{obs:bound1norm} follows. In that sense Observation~\ref{ob:quadratic_quantifier} is the stronger one.

\subsection{Hilbert-Schmidt norm}
Next, we will quantify the coherence by means of the Hilbert-Schmidt norm of the off-diagonal blocks. We make the following Observation.
\begin{observation}
Let $\varrho = \qty[\varrho_{ij}]_{i,j=0}^{N}$ be a state. Then
\begin{equation}
\label{eq:p2}
    \meas{C}_{0:1\dots N}^{HS,2}(\varrho) \equiv \sum_{i=1}^N \norm{\varrho_{0i}^{\dagger}\varrho_{0i}}_{tr} \leq  \tr\varrho_{00}(1-\tr\varrho_{00}),
\end{equation}
where equality holds, if and only if the state $\varrho$ is pure. Note that in general $\norm{X^{\dagger}X}_{tr}=\norm{X}_{HS}^2$.
\end{observation}
First, we prove the inequality. Consider the submatrices $\varrho^{(0i)}$ where the diagonal blocks are Hermitian, $\varrho_{00}^{\dagger}=\varrho_{00}$ and $\varrho_{ii}^{\dagger}=\varrho_{ii}$, and positive semidefinite, $\varrho_{00}\geq 0$ and $\varrho_{ii}\geq 0$. Then, Eq.~\eqref{eq:CSinequality} reads $\norm{\abs{\varrho_{0i}}^q}^2 \leq \norm{\varrho_{00}^q}\norm{\varrho_{ii}^q}$ for any unitarily invariant norm, with $\abs{\varrho_{0i}}=(\varrho_{0i}^{\dagger}\varrho_{0i})^{\frac{1}{2}}$. Choosing the trace norm and $q=2$, we obtain
\begin{equation}
\label{eq:inequality1}
 \sum_{i=1}^N \norm{\varrho_{0i}^{\dagger}\varrho_{0i}}_{tr} \leq \sqrt{\tr\varrho_{00}^2} \sum_{i=1}^N \sqrt{\tr\varrho_{ii}^2}.
\end{equation}
For (non-normalised) states it holds that $\tr \varrho^2\leq (\tr \varrho)^2$, with equality if and only if $\text{rk}\,\varrho = 1$, i.e. the state is pure. Hence
\begin{equation}
\label{eq:inequality2}
    \sqrt{\tr\varrho_{00}^2} \sum_{i=1}^N \sqrt{\tr\varrho_{ii}^2} \leq
    \tr\varrho_{00}\sum_{i=1}^N\tr\varrho_{ii}
    = \tr\varrho_{00}(1-\tr\varrho_{00}),
\end{equation}
which proves the inequality. Next, we prove that the inequality is saturated by pure, and pure states only. First assume that the inequality is saturated. Then it follows from Eq.~\eqref{eq:inequality2} that all the blocks must be pure, hence we can write $\varrho_{00}=\tr[\varrho_{00}]\ketbra{\hat{\varphi}_{0}}$ and $\varrho_{ii}=\tr[\varrho_{ii}]\ketbra{\hat{\varphi}_{i}}$, with $\tr\ketbra{\hat{\varphi}_{0}}=\tr\ketbra{\hat{\varphi}_{i}}=1$. Then, there exists a contraction $C_{0i}$, with $C_{0i}^{\dagger}C_{0i}\leq\id$, such that the off-diagonal blocks can be written as $\varrho_{0i}=\varrho_{00}^{1/2}C_{0i}\varrho_{ii}^{1/2}$. Using this and the previous result we can evaluate the left hand-side of Eq.~\eqref{eq:inequality2}. We obtain
\begin{eqnarray}
\norm{\varrho_{0i}^{\dagger}\varrho_{0i}}_{tr} &=& \tr[C_{0i}\varrho_{ii}C_{0i}^{\dagger}\varrho_{00}] \notag \\
&=& \tr[\varrho_{00}]\tr[\varrho_{ii}]\tr[C_{0i}\ketbra{\hat{\varphi}_{i}}C_{0i}^{\dagger}\ketbra{\hat{\varphi}_{0}}] \notag \\
&=& \tr[\varrho_{00}]\tr[\varrho_{ii}]\abs{\bra{\hat{\varphi}_{0}}C_{0i}\ket{\hat{\varphi}_{i}}}^2.
\end{eqnarray}
Then, if equality in Eq.~\eqref{eq:inequality1} holds, we must have that $\bra{\hat{\varphi}_{0}}C_{0i}\ket{\hat{\varphi}_{i}}=e^{i\varphi_{0i}}$, for some real phase $\varphi_{0i}$, and the off-diagonal blocks take the following form:
\begin{eqnarray}
    \varrho_{0i}&=&\sqrt{\tr\varrho_{00}\tr\varrho_{ii}}\ketbra{\hat{\varphi}_{0}}C_{0i}\ketbra{\hat{\varphi}_{i}}\notag \\
    &=& \sqrt{\tr\varrho_{00}\tr\varrho_{ii}}\,\,e^{i\varphi_{0i}}\ketbra{\hat{\varphi}_{0}}{\hat{\varphi}_{i}}\notag \\
    &=&\sqrt{\tr\varrho_{00}\tr\varrho_{ii}}\,\,e^{i\varphi_{0i}}\text{diag}(\varphi_0)\ketbra{+}\text{diag}(\varphi_i)^{\dagger},
\end{eqnarray}
where $\ket{+}=(1,1,\dots,1)$. Using Eq.~\eqref{eq:contraction} it also follows that all the other blocks are of rank one. We find
\begin{eqnarray}
    \varrho_{kl}&=&\sqrt{\tr\varrho_{kk}\tr\varrho_{ll}}\ketbra{\hat{\varphi}_{k}}C_{kl}\ketbra{\hat{\varphi}_{l}}\notag \\
    &=& \sqrt{\tr\varrho_{kk}\tr\varrho_{ll}}c_{kl}e^{i\varphi_{kl}}\ketbra{\hat{\varphi}_{k}}{\hat{\varphi}_{l}}\notag \\
    &=&\sqrt{\tr\varrho_{kk}\tr\varrho_{ll}}c_{kl}e^{i\varphi_{kl}}\text{diag}(\varphi_k)\ketbra{+}\text{diag}(\varphi_l)^{\dagger}.
\end{eqnarray}
So far we have shown that if the inequality is saturated, it follows that all the blocks are pure and we know the structure of the first row and the diagonal. What remains to be proven is that all the other blocks have a structure such that the overall state is pure. So far we have $\varrho = F( \tilde{\varrho}\otimes\ketbra{+})F^{\dagger}$, where $F$ is a filter (i.e. invertible) of the form $F=\text{diag}(\sqrt{\tr\varrho_{ii}}\text{diag}(\varphi_i))$. Furthermore we have that $\tilde{\varrho} = \qty[c_{ij}e^{i\varphi_{ij}}]_{i,j=0}^N$, where $c_{ii}e^{i\varphi_{ii}}=1$ and $c_{0j}=c_{i0}=1$.
Then, the state $\varrho$ must be positive semidefinite, which is the case if and only if $\tilde{\varrho}$ is positive semidefinite. Now, consider the vector $\ket{\psi}=\frac{1}{\sqrt{2}}(1-N,e^{-i\varphi_{01}},\dots, e^{-i\varphi_{0N}})$. For this vector we obtain
\begin{eqnarray}
    \bra{\psi}\varrho\ket{\psi} &=& -\frac{(N-2)(N-1)}{2}\notag \\
    &+&\sum_{1\leq i < j \leq N}c_{ij}\frac{e^{i(\varphi_{ij}-\varphi_{0j}+\varphi_{0i})}+e^{-i(\varphi_{ij}-\varphi_{0j}+\varphi_{0i})}}{2}\notag \\
    &\geq& 0.\notag
\end{eqnarray}
Since the last term in the sum is $\cos(\varphi_{ij}-(\varphi_{0j}-\varphi_{0i}))$ and there are exactly $(N-2)(N-1)/2$ terms in the sum it follows that $c_{ij}=1$ and $\varphi_{ij}=\varphi_{0j}-\varphi_{0i}$ for all $i,j$. Thus the state is of the form $\varrho = \ketbra*{\tilde{\psi}}$, with
\begin{equation}
    \label{eq:PureState}
    \ket*{\tilde{\psi}}= \bigoplus_{i=0}^N \sqrt{\tr\varrho_{ii}}\ket{\hat{\varphi}_i} e^{i\varphi_{0i}}.
\end{equation}
This proves that the global state is necessarily pure. The converse it easy to prove, since any pure state admits a decomposition in the form of Eq.~\eqref{eq:PureState}.

\subsection{Relative entropy of coherence}
Another common quantifier of coherence is the von Neumann relative entropy~\cite{baumgratz2014}. It captures the increase of entropy when going from a state $\varrho$ to its decohered version $\varrho_d$.
Here we define $\varrho^{(0i)}=\sum_{k,l=0,i} P_k\varrho P_l$ to be a trimmed version of $\varrho$ and $\varrho^{(0i)}=\sum_{k=0,i} P_k\varrho P_k$ its decohered version. We have
\begin{eqnarray}
    \sum_{i=1}^N \ent{S}(\varrho^{(0i)}|| \varrho^{(0i)}_d) = \sum_{i=1}^N \qty[\ent{S}(\varrho^{(0i)}_d)-\ent{S}(\varrho^{(0i)})].
\end{eqnarray}
For each of the terms in the sum we have
\begin{eqnarray}
\label{Eq:entropies}
    \ent{S}(\varrho^{(0i)}_d)-\ent{S}(\varrho^{(0i)}) = \tr\varrho^{(0i)}\qty[\ent{S}(\Hat{\varrho}^{(0i)}_d)-\ent{S}(\Hat{\varrho}^{(0i)})],
\end{eqnarray}
where $\hat{\varrho}=\varrho/\tr[\varrho]$. Next, we expand the first term in the sum using basic properties of the von Neumann entropy \cite{nielsen2000}. We obtain
\begin{align}
\label{eq:ineqality1}
   &\tr\varrho^{(0i)}\left[H\qty(\qty{\frac{\tr\varrho_{00}}{\tr\varrho^{(0i)}},\frac{\tr\varrho_{ii}}{\tr\varrho^{(0i)}}}) \right.\notag \\
   &\phantom{\tr\varrho^{(0i)}}+\frac{\tr\varrho_{00}}{\tr\varrho^{(0i)}}\ent{S}(\Hat{\varrho}_{00})+\left.\frac{\tr\varrho_{ii}}{\tr\varrho^{(0i)}}\ent{S}(\Hat{\varrho}_{ii}) - \ent{S}(\hat{\varrho}^{(0i)})\right] \notag \\
   &\leq \tr(\varrho^{(0i)})  h_2\qty(\frac{\tr(\varrho_{00})}{\tr(\varrho^{(0i)})}),
\end{align}
where the inequality holds since $\ent{S}(\varrho)\geq \sum_k p_k\ent{S}(\varrho_k)$, where $\varrho_k$ is the normalised post-measurement state corresponding to outcome $k$ of a projective measurement and $p_k$ are the outcome probabilities. This follows from the fact that when a projective measurement is performed and the outcome is recorded the uncertainty about the state does not increase on average, see Ref.~\cite{Fuchs2001}. Then the sum over the right-hand side of Eq.~\eqref{eq:ineqality1} can be further upper bounded by
\begin{align}
    \label{eq:ineqality2}
    &\sum_{i=1}^N \tr(\varrho^{(0i)}) h_2\qty(\frac{\tr\varrho_{00}}{\tr\varrho^{(0i)}})\notag \\
    &= \sum_{k=1}^N \tr\varrho^{(0k)} \sum_{i=1}^N \frac{\tr\varrho^{(0i)}}{\sum_{k=1}^N \tr\varrho^{(0k)}} h_2\qty(\frac{\tr\varrho_{00}}{\tr\varrho^{(0i)}})\notag \\
   &\leq \qty[(N-1)\tr\varrho_{00}+1]h_2\qty(\frac{N\tr\varrho_{00}}{(N-1)\tr\varrho_{00}+1}),
\end{align}
where we have used that $\sum_{k=1}^N \tr\varrho^{(0k)}=(N-1)\tr \varrho_{00}+1$ and the concavity of the binary entropy, namely that $\sum_i p_i h_2(x_i)\leq h(\sum_i p_ix_i)$.
We arrive at the following observation.
\begin{observation}\label{obs:entropictradeoff}
The amount of coherence that is shared between the first and all other subspaces, quantified by the relative entropy of block coherence, is bounded by
\begin{align}\label{eq:qutritentropy}
    &\sum_{i=1}^N\ent{S}(\varrho^{(0i)}|| \varrho^{(0i)}_d)\notag \\
    &\leq \qty[(N-1)\tr(\varrho_{00})+1]h_2\qty(\frac{N\tr(\varrho_{00})}{(N-1)\tr(\varrho_{00})+1}).
\end{align}
\end{observation}
The bound in this inequality is also saturated by the state from Eq.~\eqref{eq:saturatingstate}. In the first inequality in Eq.~\eqref{eq:ineqality1} equality holds because $\Hat{\sigma}_{00}$ and $\Hat{\sigma}_{ii}$ have the same entropy as $\sigma$. Note, that we have $\sigma=\ketbra{\varphi}\otimes \Hat{\varrho}_{00}$, $\sigma_{00}=(\ketbra{0}\otimes \id)\sigma (\ketbra{0}\otimes \id)$ as well as $\sigma_{ii}=(\ketbra{i}\otimes \id)\sigma (\ketbra{i}\otimes \id)$. In the second inequality on Eq.~\eqref{eq:ineqality2} $\tr(\sigma_{00})/\tr(\sigma^{0i})$ is the same for all $i=1,\dots,N$, and hence equality holds due to the concavity of the binary entropy $h_2$.

\section{Example of a single qutrit} \label{sec:4}

\begin{figure}[t!]
    \centering
    \includegraphics[scale=.7]{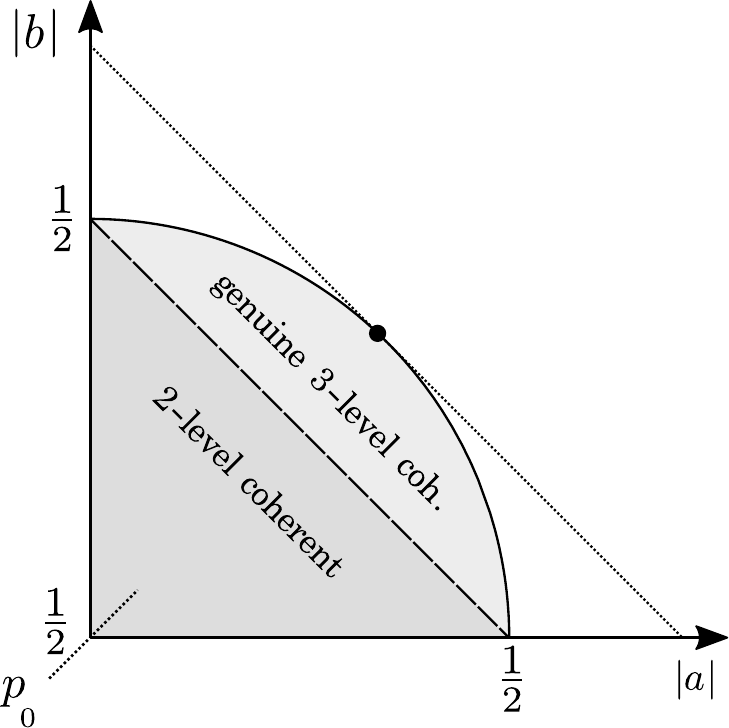}
\caption{This figure shows the trade-off relations for a single qutrit state. The coherences between any two subspaces $\abs{a}$ and $\abs{b}$ are bounded by $\frac{1}{2}$. The bound from Observation~\ref{obs:bound1norm} (dotted line) is tight, but does not completely characterize the set of physical realizable states (grey area). The quantifier from Observation~\ref{ob:quadratic_quantifier} is also tight and it completely characterizes the set of physical realizable states (solid line). The dashed line separates the two-level coherent states from genuine three-level coherent states.}
\label{fig:qutrit1norm}
\end{figure}

Let us now discuss the results of the previous section in the case of a single qutrit and $N=2$. We write $\varrho$ as
\begin{equation}
\label{eq:qutrit_example}
    \varrho = \mqty(p_0 & a & b \\ \bar{a} & p_1 & \cdot \\ \bar{b} & \cdot & p_2),
\end{equation}
$p_2=1-p_0-p_1$, where we have used dots for placeholders for entries that we do not directly consider. For a given value of $p_0$ the physical region is bounded by the inequalities
\begin{equation}
    \abs{a} + \abs{b} \leq \sqrt{2}\sqrt{p_0(1-p_0)}
\end{equation}
and
\begin{equation}
\label{eq:qudraticqutrit}
    \abs{a}^2 + \abs{b}^2 \leq p_0(1-p_0),
\end{equation}
which are special cases of Observation \ref{obs:bound1norm} and Observation \ref{ob:quadratic_quantifier}, respectively.
The physical region, see Figure \ref{fig:qutrit1norm}, corresponds to a quarter disk, where all pure states lie on the border defined by the quarter circle. The states lying on the $a$ and $b$ axis are two-level coherent states. By varying the parameter $p_0$ one sees, that the set of physical states is given by a quarter ball, where all pure states lie on the quarter sphere.

For the entropic trade-off relation in Observation~\ref{obs:entropictradeoff} we find the following. For $N=2$ the two entropies $\ent{S}(\varrho^{(01)}|| \varrho^{(01)}_d)=\ent{S}_{01}$ and $\ent{S}(\varrho^{(02)}|| \varrho^{(02)}_d)=\ent{S}_{02}$ are both bounded by $1$. For their sum we obtain the bound
\begin{equation}
    \ent{S}_{01} + \ent{S}_{02} \leq (p_0+1)h_2\qty(\frac{2p_0}{p_0+1}).
\end{equation}
Computing the maximum value for this bound by maximizing over $p_0$ one obtains $\ent{S}_{01} + \ent{S}_{02} \leq \qty(1-\frac{2}{\sqrt{5}})+ \frac{1-\sqrt{5}}{\sqrt{5}}\log_2(3-\sqrt{5})+\frac{1}{\sqrt{5}}\log_2(3+\sqrt{5})\approx 1.3885$, for $p_0=\frac{1}{\sqrt{5}}$.

\section{Detection of genuine multisubspace coherence}\label{sec:5}
Any pure state $\ket{\psi}$ can be written as $\ket{\psi} = \sum_{j=0}^N \ket{\psi_j}$, where $\ket{\psi_j}=P_j \ket{\psi}$. We say that $\ket{\psi}$ has block coherence rank (BCR) equal to $k$ if exactly $k$ of the $\ket{\psi_i}$ do not vanish. We denote $BC_k$ the convex hull of all pure states with block coherence at most $k$. We say that a mixed state $\rho$ has block-coherence number $BCN(\rho)$ equal to $k$ if it is in $BC_k$ but not in $BC_{k-1}$. We are going to see that the quantifier $\meas{C}^{tr}_{0:1\dots N}$ obeys a stricter bound for states with limited block coherence.

\begin{observation}
It holds
\begin{equation}
    \label{eq:BCNbound}
    \meas{C}^{tr}_{0:1\dots N} (\rho)\leq \sqrt{BCN(\varrho)-1} \sqrt{{\tr\varrho_{00}(1-\tr\varrho_{00})}}.
\end{equation}
\end{observation}
If a pure state $\ket{\psi}$ has block-coherence rank $BCR(\ket{\psi})$, then
\begin{equation}
\label{eq:boundpurerank}
\begin{split}  
\meas{C}^{tr}_{0:1\dots N}(\ketbra{\psi}{\psi})&\leq \sqrt{BCR(\ket{\psi})-1}\\  &\quad\times\sqrt{\braket{\psi_0}(1-\braket{\psi_0}})
\end{split}
\end{equation} 
since there are at most $BCR(\ket{\psi})-1$ other blocks that are populated. A state $\rho$ admits a pure-state ensemble decomposition $\rho = \sum_j p_j \ketbra*{\psi^{j}}$ with $BCR(\ket*{\psi^{(j)}})\leq BCN(\rho)$. Thus,
\begin{equation}
\begin{split}
    \meas{C}^{tr}_{0:1\dots N} (\rho)
    &\leq \sum_j p_j \meas{C}_{0:1\dots N} (\ketbra*{\psi^{j}}) \\
    &\leq \sqrt{BCN(\rho)-1}\\
    &\quad\times\sum_j p_j \sqrt{\braket*{\psi^{j}_0}(1-\braket*{\psi^{j}_0}}) \\
    &\leq \sqrt{BCN(\rho)-1}\\
    &\quad\times\sqrt{{\sum_j p_j\braket*{\psi^{j}_0} (1-\sum_j p_j\braket*{\psi^{j}_0})}}\\
    &= \sqrt{BCN(\rho)-1} \sqrt{{\tr\varrho_{00}(1-\tr\varrho_{00})}}.
\end{split}
\end{equation} 
The first inequality is due to the convexity of $\meas{C}^{tr}_{0:1\dots N}$, the second inequality is due to the bound in Eq.~\eqref{eq:boundpurerank}, and the third inequality is due to the concavity in $0\leq x\leq 1$ of $\sqrt{x(1-x)}$.

Notice that recover the always valid bound in Eq.~\eqref{eq:bound1norm} by considering that $BCN(\rho)\leq N+1$. One benefit of the bound in Eq.~\eqref{eq:BCNbound} is that it allows one to certify genuine multi-block coherence just by considering one block-row of the density matrix.

For the qutrit example in Eq.~\eqref{eq:qutrit_example} of Section~\ref{sec:4} the inequality in Eq.~\eqref{eq:BCNbound} reads explicitly $\meas{C}^{tr}_{0:1\dots N} (\rho)\leq \sqrt{{\tr\varrho_{00}(1-\tr\varrho_{00})}}$ for any state with block-coherence number less or equal to two. Hence, any state of the form in Eq.~\eqref{eq:qutrit_example} that violates this necessarily contains three-level coherence. See Figure~\ref{fig:qutrit1norm} for $p_0=1/2$, i.e., $\abs{a}+\abs{b}\leq 1/2$.

\section{Conclusions}
In this manuscript we have derived trade-off relations for the coherence that can be shared between multiple subspaces, as a consequence of the positivity constraint on the density matrix. We formulated trade-off relations in terms of the trace norm and Hilbert-Schmidt norm as well as the von Neumann relative entropy. Furthermore, we found that out quantifier can be used to detect multisubspace coherence, i.e., that it obeys stricter bound when states with limited block coherence are considered. We further conclude that similar trade-offs also hold for other positive semidefinite matrices, e.g., covariance matrices and Choi matrices.
For future research it would be interesting to see the implications of this trade-off in applications in which block coherence is important, e.g., in quantum clocks~\cite{Hyukjoon2018} or quantum metrology. Furthermore it would be interesting to study the connection to entanglement monogamy and to see to what extent this trade-off is relevant for genuine multipartite entanglement.

\begin{acknowledgments}
We thank O. Gühne for discussions and useful comments on the manuscript.  This work was supported by the DFG and the ERC (Consolidator Grant 683107/TempoQ).
\end{acknowledgments}

\bibliography{bibliography.bib}

\end{document}